\documentclass[twocolumn,groupedaddress]{revtex4-1}
\usepackage{amsmath, bm, physics}
\usepackage{color}
\usepackage{graphicx}
\usepackage{siunitx}

\usepackage[whole]{bxcjkjatype} 
\usepackage[colorlinks=true,urlcolor=blue,citecolor=blue,linkcolor=blue,breaklinks=true]{hyperref}
\input glyphtounicode
\usepackage{txfonts}

\newcommand{\smrm}[1]{_{\mathrm{#1}}}

\begin{document}
\title{Time-dependent Gutzwiller simulation of Floquet topological superconductivity}
\author{Takahiro Anan}
\affiliation{Department of Applied Physics, The University of Tokyo, Hongo, Tokyo,
113-8656, Japan}
\author{Takahiro Morimoto}
\affiliation{Department of Applied Physics, The University of Tokyo, Hongo, Tokyo,
113-8656, Japan}
\author{Sota Kitamura}
\affiliation{Department of Applied Physics, The University of Tokyo, Hongo, Tokyo,
113-8656, Japan}
\date{\today}
\begin{abstract}
Periodically driven systems provide a novel route to control the topology of quantum materials. In particular, Floquet theory allows an effective band description of periodically-driven systems through the Floquet Hamiltonian. Here, we study the time evolution of $d$-wave superconductors irradiated with intense circularly-polarized laser light. We consider the Floquet $t$--$J$ model with time-periodic interactions, and investigate its mean-field dynamics by formulating the time-dependent Gutzwiller approximation. We observe the development of the $id_{xy}$-wave pairing amplitude along with the original $d_{x^2-y^2}$-wave order upon gradual increasing of the field amplitude. We further numerically construct the Floquet Hamiltonian for the steady state, with which we identify the system as the fully-gapped $d+id$ superconducting phase with a nonzero Chern number. We explore the low-frequency regime where the perturbative approaches in the previous studies break down, and find that the topological gap of an experimentally-accessible size can be achieved at much lower laser intensities.
\end{abstract}
\maketitle

\section*{Introduction}
Topological superconductors host robust gapless excitations at the boundary or vortex cores due to the topological structure of the superconducting gap function~\cite{Kitaev2001,Read2000,Sato2016,Chiu-RMP16}. In particular, Majorana fermions that emerge in topological superconductors provide a platform for fault-tolerant quantum computation~\cite{Kitaev2003}.
Theoretical proposals for creating a topological superconductor include topological insulators in the proximity of $s$-wave superconductors~\cite{Fu2008} and semiconductors with spin-orbit coupling in the proximity of $s$-wave superconductors~\cite{Lutchyn2010}.
Despite intense experimental efforts to confirm topological superconductivity and Majorana fermions in those setups, their existence is still elusive~\cite{Yazdani2023}. Hence, seeking an alternative platform for topological superconductivity remains an important issue.

Periodically driven systems provide a novel route to control the topology of quantum materials. In particular, Floquet theory allows an effective band description of periodically-driven systems through the Floquet Hamiltonian. Thus the dynamical control of quantum phases has recently been studied actively, called ``Floquet engineering''~\cite{Oka2019, Bukov2015,Eckardt2017,Torre2021,Morimoto2023}. A canonical example of Flouqet engineering of a topological phase is the quantum anomalous Hall state that emerges in graphene irradiated by circularly-polarized light (CPL)~\cite{Oka2009,Kitagawa2011,Jotzu2014,McIver2020}.
In graphene, CPL induces an effective complex hopping for the next-nearest neighbor in the Floquet Hamiltonian, which takes the same form as in the Haldane model for the quantum anomalous Hall state~\cite{Haldane1988}. 

Applying the concept of Floquet engineering to topological superconductors, Floquet topological superconductivity has been explored~\cite{Ezawa2015,Zhang2015,Takasan2017,Dehghani2017,Chono2020,Kumar2021,Dehghani2021}.
For example, a honeycomb lattice with $s$-wave pairing interaction is predicted to exhibit topological superconductivity under the irradiation of CPL \cite{Ezawa2015,Chono2020}, where CPL induces a mass term to gapless excitations around point nodes, leading to topological superconductivity.
A similar strategy for Floquet topological superconductivity was also pursued for cuprate superconductors. Specifically, a $d$-wave superconductor on a square lattice was shown to support CPL-induced topological superconductivity when strong spin-orbit coupling (SOC) is present~\cite{Takasan2017}. All these approaches to Floquet topological superconductivity essentially rely on the presence of additional internal degrees of freedom supporting nontrivial geometry (i.e. sublattice in the honeycomb lattice or spins with SOC) for turning the systems into topological phases.

\begin{figure}[t]
\begin{center}
\includegraphics[width=\linewidth]{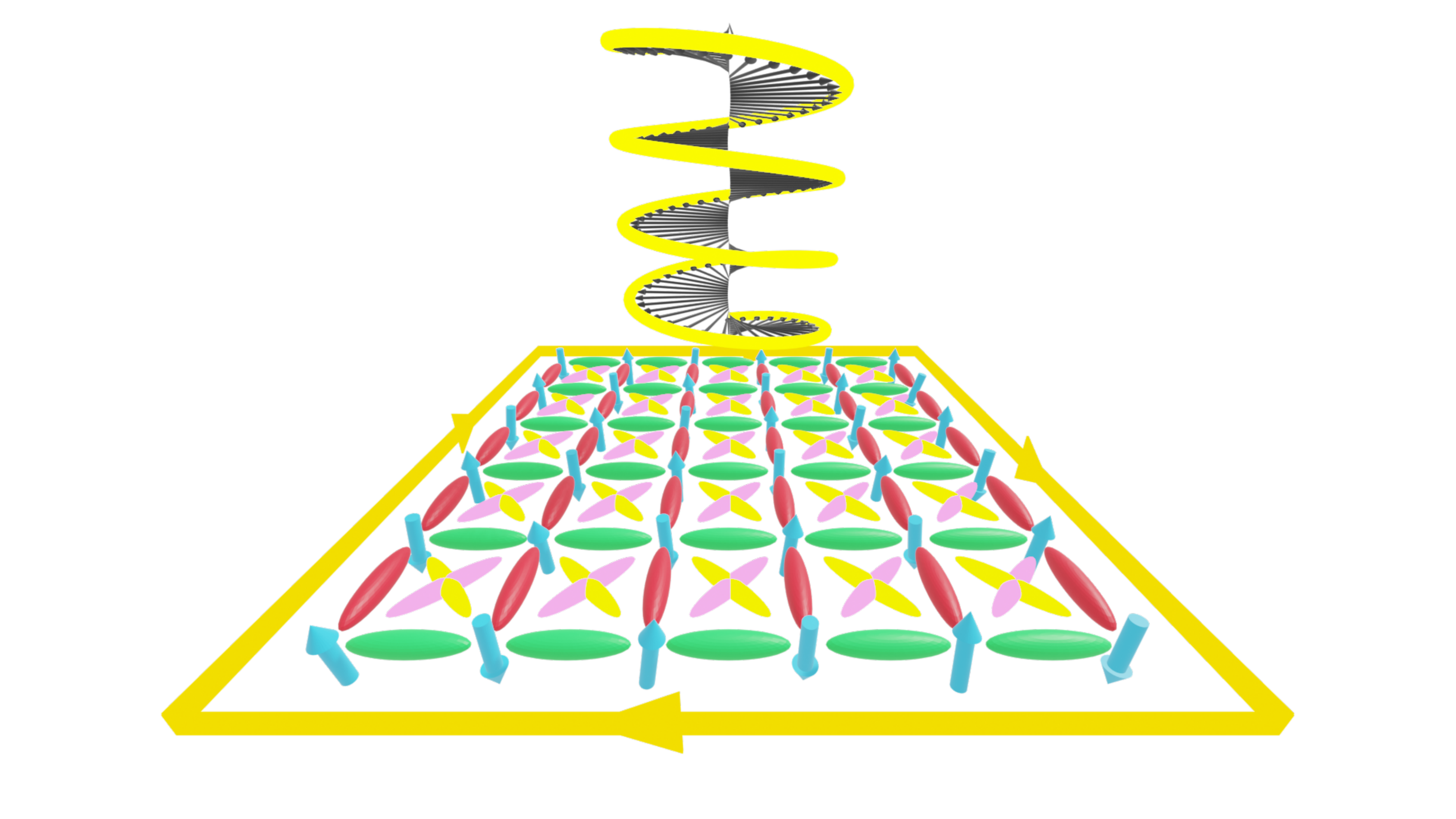}
\caption{Schematics of Floquet topological superconductivity.
Illuminating circularly polarized light to cuprate superconductors gives rise to topological superconductivity of $d_{x^2-y^2}+id_{xy}$-wave pairing from many body effects.
}
\label{fig:1}
\end{center}
\end{figure}

Recently, it was revealed that $d$-wave superconductors exhibit Floquet topological superconductivity purely from the many-body effect without invoking the internal degrees of freedom~\cite{Kitamura2022}. To incorporate strong correlation effects, they derive the Floquet $t$--$J$ Hamiltonian using the Schrieffer-Wolff transformation~\cite{Mentink2014,Bukov2016,Kitamura2017,Claassen2017} and the high-frequency expansion (HFE)~\cite{Bukov2015,Eckardt2015,Mikami2016}. Time-reversal symmetry breaking terms appear from the interaction terms in the Floquet $t$--$J$ model and induce topological $d_{x^2-y^2}+id_{xy}$ pairing upon a mean field treatment (See Fig.~\ref{fig:1}). This approach enables us to broaden the class of candidate materials for the Floquet topological superconductivity.

However, experimental implementation of the above theoretical proposals is still challenging because of the required field intensity of $\sim\SI{100}{\mega\V\per\cm}$~\cite{Takasan2017,Kitamura2022}. 
This stringent requirement essentially stems from the fact that the HFE has been employed in these previous studies. The field-induced effective coupling in the HFE typically scales with the amplitude of the vector potential, $A\propto E/\omega$, which implies that a strong electric field $E$ is necessary for a large driving frequency $\omega$.
Also, driving in the low frequency regime below the electronic band gap is desirable to avoid heating of the system and achieve a coherent control. 
Thus it is essential to develop a theoretical framework that is applicable to the low frequency regime.

In this paper, we study Floquet topological superconductivity at low frequencies by performing a time-dependent Gutzwiller simulation.
We formulate the time-dependent Gutzwiller approximation based on the action principle combined with the mean-field approximation.
Specifically, to simulate the time evolution of the many-body state in a tractable way, we apply the Gutzwiller approximation to the Lagrangian formalism to deduce a time-dependent Schr\"{o}dinger equation with a mean field approximation for pairing amplitudes and bond orders. The time-dependent simulation shows that the $d_{x^2-y^2}$-wave superconductor evolves into the topological $d_{x^2-y^2}+id_{xy}$-wave superconductor under the CPL driving, both in the high and low frequency regimes. 
We further analyze the obtained time-periodic superconducting state in terms of the Floquet Hamiltonian, revealing the full-gap nature and the nontrivial winding of the gap function.
We find that the topological gap of the order of \SI{3}{\K} emerges for an electric field of $\sim \SI{10}{\mega\V\per\cm}$, which will be feasible for experimental measurements.

\section*{Results}
\subsection*{Formalism}\label{sec:2}
In this section, we derive an effective low-energy Bogoliubov-de Gennes (BdG) Hamiltonian in the presence of the CPL, employing time-periodic Schrieffer-Wolff transformation (a canonical transformation) with Gutzwiller ansatz~\cite{Kitamura2022}. 

We consider a periodically-driven Hubbard model defined on a square lattice, having in mind a cuprate superconductor.
The time-dependent Hamiltonian is given by
\begin{align}
    \hat{H}_\mathrm{Hub}(t)&=-\sum_{ij\sigma }t_{ij}e^ {-i\bm{A}(t)\cdot \bm{R}_{ij}}\hat{c}_{i\sigma}^\dagger \hat{c}_{j\sigma}+U\sum_i \hat{n}_{i\uparrow} \hat{n}_{i\downarrow},
\end{align}
where $\hat{c}_{i\sigma}$ is an electron annihilation operator at site $i$ with spin $\sigma=\uparrow,\ \downarrow$, and $\hat{n}_{i\sigma } = \hat{c}_{i\sigma}^\dagger \hat{c}_{i\sigma}$ is the spin-density operator.
Here we set $\hbar =e=1$ for simplicity.
The latter term is the on-site Coulomb repulsion terms with the Hubbard interaction $U$, while 
the former term is the hopping terms with modulated hopping amplitude $t_{ij}e^ {-i\bm{A}(t)\cdot \bm{R}_{ij}}$, where $t_{ij}$ is the hopping amplitude between site $i$ and site $j$. We introduce $\bm{R}_{ij}=\bm{R} _{i} -\bm{R}_j$, where $\bm{R}_i$ is the location of the $i$-th site.
Here we consider CPL, for which the vector potential $\bm{A}(t)$ is given by
\begin{align}
    \bm{A}(t) &=\frac{1}{2}(\bm{A}_0 e^{-i\omega t}+\bm{A}_0^* e^{i\omega t}),\\
    &\ \ \ \bm{A}_0=\frac{E}{i\omega }\left(\begin{array}{c}
        1 \\
        i 
    \end{array}\right).
\end{align}
To deduce low-energy dynamics of the driven Hubbard model, we consider the Lagrangian of this system,
\begin{align}
  L=\bra{\Psi(t)}(i\partial_t - \hat{H}_\mathrm{Hub}(t))\ket{\Psi(t)},
\end{align}
where $\ket{\Psi(t)}$ is a state vector of the many-body system.

\begin{figure}[t]
  \begin{center}
  \includegraphics[width=0.6\linewidth]{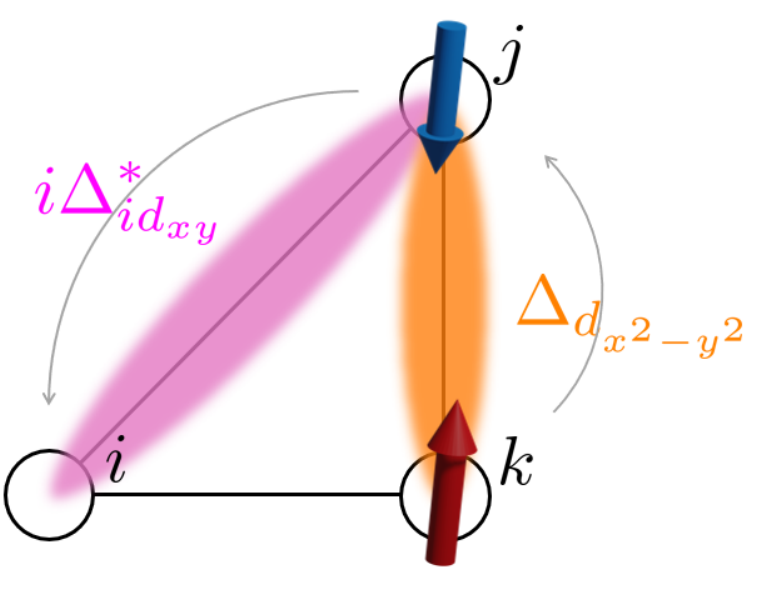}
  \caption{Schematic picture of the $id_{xy}$-wave component of the pairing amplitude $\Delta_{id_{xy}}$ that arises from three-site terms.
  Through the virtual hopping process from the site $k$ to the site $i$ via the site $j$, the  $id_{xy}$-wave component $\Delta_{id_{xy}}$ (magenta) is induced from the original $d_{x^2-y^2}$-wave order (orange).}
  \label{fig:Sup:threeSiteTerm}
\end{center}
\end{figure}

First, we perform the time-periodic Schrieffer-Wolff transformation~\cite{Bukov2016,Kitamura2017},
where the transformed state vector is represented as $\hat{P}_G e^{i\hat{S}(t)}\ket{\Psi(t)}$, with the unitary transformation 
$e^{i\hat{S}(t)}$ and the Gutzwiller projection $\hat{P}_G=\prod _i(1-\hat{n}_{i\uparrow} \hat{n}_{i\downarrow})$.
Here $\hat{S}(t)$ should be chosen such that the transformed Hamiltonian,
$\hat{H}_\mathrm{SW}(t) \equiv \hat{P}_G(e^{i\hat{S}(t)}\hat{H}_\mathrm{Hub}e^{-i\hat{S}(t)}-e^{i\hat{S}(t)}(i\partial_te^{-i\hat{S}(t)}))\hat{P}_G$, becomes diagonal in the charge sector (eliminating charge excitations), and thus commutes with $\hat{P}_G$ (For a similar method for the Hubbard model not based on the Schrieffer-Wolff transformation but with a generalized projection operator, see Refs.~\cite{Schiro2010,Schiro2011}).
Then the transformed state vector only takes the configurations that have no doubly-occupied site.
By adopting the Gutzwiller ansatz where $e^{i\hat{S}(t)}\ket{\Psi(t)}$ is chosen to be the BCS wave function $\ket{\Psi_\mathrm{BCS}(t)}=\prod_{\bm{k}}(u_{\bm{k}}(t)+v_{\bm{k}}(t)\hat{c}_{\bm{k}\uparrow}^\dagger\hat{c}_{-\bm{k}\downarrow}^\dagger)\ket{0}$,
here we approximate the original Lagrangian $L$ by $L_G$, as
\begin{align}
  L_G=&\frac{\bra{\Psi(t)}e^{-i\hat{S}(t)}\hat{P}_Ge^{i\hat{S}(t)}(i\partial_t - \hat{H}_\mathrm{Hub}(t))e^{-i\hat{S}(t)}\hat{P}_Ge^{i\hat{S}(t)}\ket{\Psi(t)}}{\bra{\Psi(t)}e^{-i\hat{S}(t)}\hat{P}_G\hat{P}_Ge^{i\hat{S}(t)}\ket{\Psi(t)}}\nonumber \\
=&\frac{\bra{\Psi_\mathrm{BCS}(t)}\hat{P}_Gi\partial_t \hat{P}_G\ket{\Psi_\mathrm{BCS}(t)}}{\bra{\Psi_\mathrm{BCS}(t)}\hat{P}_G\hat{P}_G\ket{\Psi_\mathrm{BCS}(t)}}\nonumber \\&- \frac{\bra{\Psi_\mathrm{BCS}(t)}\hat{P}_G\hat{H}_\mathrm{SW}(t)\hat{P}_G\ket{\Psi_\mathrm{BCS}(t)}}{\bra{\Psi_\mathrm{BCS}(t)}\hat{P}_G\hat{P}_G\ket{\Psi_\mathrm{BCS}(t)}}. \label{LG}
\end{align} 
We conduct the Schrieffer-Wolff transformation up to the second-order of the hopping and obtain~\cite{Kitamura2022}
\begin{align}
  \hat{H}\smrm{SW}(t) =&-\sum_{ij\sigma}\tilde{t}_{ij}(t)\hat{P}_G\hat{c}_{i\sigma}^\dagger \hat{c}_{j\sigma}\hat{P}_G \nonumber \\
  &+\frac{1}{2}\sum_{ij}\tilde{J}_{ij}(t)\hat{P}_G \left[\hat{\bm{S}}_i \cdot \hat{\bm{S}}_j-\frac{1}{4}\hat{n}_i\hat{n}_j\right]\hat{P}_G  \nonumber\\
    &+\sum_{ijk\sigma \sigma'}^{i\neq k}\Bigg\{\tilde{\Gamma}_{ijk}(t)\hat{P}_G\Bigg[(\hat{c}_{i\sigma}^\dagger \bm{\sigma}_{\sigma \sigma '}\hat{c}_{k\sigma '})\cdot \hat{\bm{S}}_j\nonumber \\
    &\hspace{2.8cm} -\frac{1}{2}\delta_{\sigma \sigma '}\hat{c}_{i\sigma}^\dagger \hat{c}_{k\sigma}\hat{n}_j\Bigg]\hat{P}_G+h.c. \Bigg\}    ,\label{HSW}
\end{align}
where $\hat{\bm{S}}_j$ is a spin operator. 
The coupling constants are given by
\begin{align}
  \tilde{t}_{ij}(t)&=\sum_{m}t_{ij}^{(m)}e^{-im\omega t},\\
  \tilde{J}_{ij}(t)&=\sum_{mn}\frac{4e^{-im\omega t}}{U-n\omega }t_{ij}^{(m-n)}t_{ji}^{(n)},\\
  \tilde{\Gamma}_{ijk}(t)&=\sum_{mn}\frac{e^{-im\omega t}}{2(U-n\omega )}t_{ij}^{(m-n)}t_{jk}^{(n)}.
\end{align}
Here, $t_{ij}^{(m)}$ is the $m$-th Fourier component of the modulated hopping amplitude $t_{i j} e^{-i \bm{A}(t) \cdot \bm{R}_{i j}}$, whose explicit form is obtained by the Jacobi-Anger expansion as
\begin{align}
    t_{i j}^{(m)}&=\frac{\omega}{2 \pi} \int_0^{2 \pi / \omega} d t\ t_{i j} e^{-i \bm{A}(t) \cdot \bm{R}_{i j}+i m \omega t}  
      =t_{i j} \mathcal{J}_{-m}\left(\frac{E|\bm{R}_{ij}|}{\omega}\right) e^{i m \Theta_{i j}} \label{tijm}
\end{align}
with the $m$-th Bessel function $\mathcal{J}_m(x)$  and $\Theta _{ij}$ defined as the polar angle of $\bm{R}_{ij}$, i.e., $\bm{R}_{ij}=|\bm{R}_{ij}|(\cos \Theta _{ij},\sin \Theta _{ij})$. 
In the absence of the external field, Eq.~\eqref{HSW} is known as the $t$--$J$ model~\cite{Ogata2008}.
The first term is the hole hopping term in the configurations that have no doubly-occupied site.
The second term is composed of Heisenberg interaction $\hat{\bm{S}}_i \cdot \hat{\bm{S}}_j$ and the density-density interaction $\hat{n}_i\hat{n}_j$.
The third term, representing an interaction of the form $\sum_{\sigma \sigma'}[(\hat{c}_{i\sigma}^\dagger \bm{\sigma}_{\sigma \sigma '}\hat{c}_{k\sigma '})\cdot \hat{\bm{S}}_j-\frac{1}{2}\delta_{\sigma \sigma '}\hat{c}_{i\sigma}^\dagger \hat{c}_{k\sigma}n_j]$, is the so-called three-site term.
This Hamiltonian is known to yield singlet ($d$-wave) Cooper pairing in equilibrium, so that here we assume the singlet pairing as well.
A remarkable point here is that the three-site terms in the present case break time-reversal symmetry, due to the circular polarization of the light field. 
When we consider the spin part of the three-site terms $\sum_{\sigma \sigma'}\hat{c}_{i\sigma}^\dagger \bm{\sigma}_{\sigma \sigma'} \hat{c}_{k \sigma'}\cdot  \hat{\bm{S}}_j$ where $i$-$j$ is a next-nearest neighbor bond and $j$-$k$ is a nearest neighbor bond, the three-site terms generate pairing amplitude on each bond as $\sum_{\sigma \sigma'}\hat{c}_{i\sigma}^\dagger \bm{\sigma}_{\sigma \sigma'} \hat{c}_{k \sigma'}\cdot \hat{\bm{S}}_j \sim \ev{\hat{c}_{i\downarrow}^\dagger\hat{c}_{j\uparrow}^\dagger}\ev{\hat{c}_{j\uparrow}\hat{c}_{k\downarrow}}$. 
 The pairing amplitudes $\ev{\hat{c}_{i\downarrow}^\dagger\hat{c}_{j\uparrow}^\dagger}$ and 
$\ev{\hat{c}_{j\uparrow}\hat{c}_{k\downarrow}}$ correspond to $d_{xy}$-wave components and $d_{x^2-y^2}$-wave components of pairing amplitudes respectively.
When the coefficients of the three-site terms $\tilde{\Gamma}(t)$ in Eq.~\eqref{HSW} become complex due to the CPL, the next-nearest neighbor pairing amplitude $\ev{\hat{c}_{i\downarrow}^\dagger\hat{c}_{j\uparrow}^\dagger}$ becomes complex, embodying the $id_{xy}$-wave component of the pairing amplitude (as shown in Fig.~\ref{fig:Sup:threeSiteTerm}).
The same applies to the density part of the three-site term $\sum_{\sigma}\hat{c}_{i\sigma}^\dagger\hat{c}_{k\sigma}\hat{n}_j$.
These terms have already been shown to induce topological superconductivity by performing the HFE of Eq.~\eqref{HSW} in Ref.~\cite{Kitamura2022}.
The time-periodic Schrieffer-Wolff expansion is a perturbative expansion in condition $t_0 \ll U$ meanwhile there is no resonance between $\omega$ and $U$ (i.e. $t_{ij}^{(m)} \ll U-n\omega$).
For the detailed derivation, see Schrieffer-Wolff transformation subsection in Methods.

\begin{figure*}[t]
\begin{center}
\includegraphics[width=\linewidth]{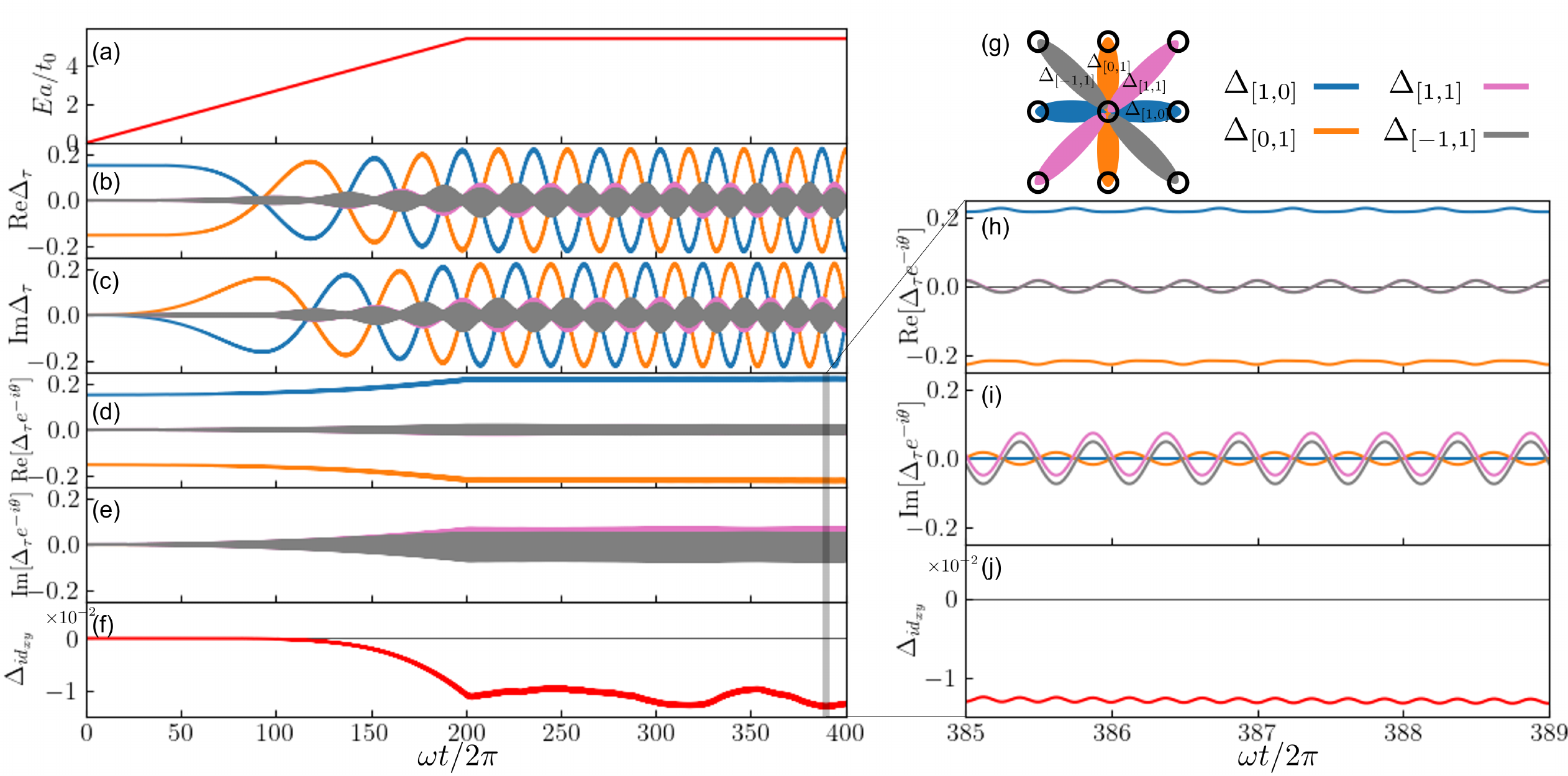}
\caption{Time evolution of superconducting pairing amplitudes obtained from time-dependent Gutzwiller simulation. 
(a) Time profile of the driving field of CPL. The electric field increases linearly up to $E=5.4t_0/a$ until $\omega t/2\pi=200$ and is constant after that.
(b, c) real and imaginary parts of the pairing amplitudes $\Delta_{\tau}$. 
(d,e) real and imaginary parts of the pairing amplitudes with the phase rotation $\Delta_{\tau}e^{-i\theta}$.
$\theta$ is selected to keep $\Delta_{[1,0]}e^{-i\theta}$ to be real (i.e., $\Delta_{[1,0]}=|\Delta_{[1,0]}|e^{i\theta}$).
(f) The $id_{xy}$-wave component of the pairing amplitude $\Delta_{id_{xy}}$. $\Delta_{id_{xy}}=(\mathrm{Im}[\Delta_{[-1,1]}e^{-i\theta}]-\mathrm{Im}[\Delta_{[1,1]}e^{-i\theta}])/2$ corresponds to the $id_{xy}$ pairing amplitude. $\Delta_{id_{xy}}$ emerges under the CPL driving and the topological superconductivity is realized.
(h,i,j) Blowup of shaded areas in (d,e,f). 
$\mathrm{Im}[\Delta_{[-1,1]}e^{-i\theta}]$ and $\mathrm{Im}[\Delta_{[1,1]}e^{-i\theta}]$ become different and nonzero $\Delta_{id_{xy}}$ appears in the steady state under the CPL driving. 
(g) Notations for the superconducting pairing amplitudes on the bonds and their color codes for (b-e,h,i).
We adopted the parameter set: 
the driving frequency $\omega=5.8t_0$,
the next nearest neighbor hopping $t_0'=-0.2t_0$, the onsite interaction $U=12t_0$ and the hole doping level $\delta=0.2$.
}
\label{fig:2}
\end{center}
\end{figure*}

Second, we perform the Gutzwiller approximation~\cite{Ogata2003,Ogawa1975,Schiro2010,Schiro2011}, which replaces the Gutzwiller projection $\hat{P}_G$ with c-numbers renormalizing each term in the Hamiltonian and effectively incorporates the reduction of double occupancies.
We perform the Gutzwiller approximation to the second term of Eq.~\eqref{LG} and derive an effective Hamiltonian $\hat{H}_G$ as 
\begin{align}
  \frac{\bra{\Psi_\mathrm{BCS}(t)}\hat{P}_G\hat{H}_\mathrm{SW}(t)\hat{P}_G\ket{\Psi_\mathrm{BCS}(t)}}{\bra{\Psi_\mathrm{BCS}(t)}\hat{P}_G\hat{P}_G\ket{\Psi_\mathrm{BCS}(t)}}&\simeq \bra{\Psi_\mathrm{BCS}(t)}\hat{H}\smrm{G}(t)\ket{\Psi_\mathrm{BCS}(t)}
  \nonumber\\&\equiv \ev{\hat{H}_G}. \label{HSWtoHG}
\end{align}
Hereafter, $\ev{\cdots}$ represents the expectation value in terms of BCS wavefunction $\ket{\Psi_\mathrm{BCS}(t)}$.
As described in Gutzwiller approximation subsection in the Methods, we obtain the effective Hamiltonian $\hat{H}_G(t)$,
\begin{align}
  \hat{H}\smrm{G}(t) =&-\sum_{ij\sigma}\frac{\delta}{\bar{f}}\tilde{t}_{ij}(t)\hat{c}_{i\sigma}^\dagger \hat{c}_{j\sigma}-\left[
  \sum_{ijk\sigma}^{i\neq k}\frac{\delta f^2}{\bar{f}^2}\tilde{\Gamma}_{ijk}(t)\hat{c}_{i\sigma}^\dagger \hat{c}_{k\sigma}+h.c.\right] \nonumber \\
  &+\frac{1}{2}\sum_{ij}\frac{1}{\bar{f}^2}\tilde{J}_{ij} (t)\left[\hat{\bm{S}}_i \cdot \hat{\bm{S}}_j -\frac{1}{4}\delta^2\hat{n}_i \hat{n}_j\right]\nonumber \\
  &+\Bigg\{\sum_{ijk\sigma \sigma'}\frac{\delta}{\bar{f}^2}\tilde{\Gamma}_{ijk}(t)\bigg[(\hat{c}_{i\sigma}^\dagger \bm{\sigma}_{\sigma \sigma '}\hat{c}_{k\sigma '})\cdot \bm{\hat{S}}_j \nonumber\\
  &\hspace{2.8cm}-\frac{1}{2}\delta \delta_{\sigma \sigma '}\hat{c}_{i\sigma}^\dagger \hat{c}_{k\sigma}\hat{n}_j\bigg]+h.c.  \Bigg\}, \label{HG}
\end{align}
where $\delta=1-\frac{1}{N}\sum_i \ev{\hat{n}_{i\uparrow}+\hat{n}_{i\downarrow}}$ is the hole doping rate, $f=(1-\delta)/2$, and $\bar{f}=1-f$. 
Similarly, we perform Gutzwiller approximation to the first term of Eq.~\eqref{LG} and obtain
\begin{align}
  \frac{\bra{\Psi_\mathrm{BCS}(t)}\hat{P}_Gi\partial_t \hat{P}_G\ket{\Psi_\mathrm{BCS}(t)}}{\bra{\Psi_\mathrm{BCS}(t)}\hat{P}_G\hat{P}_G\ket{\Psi_\mathrm{BCS}(t)}}\simeq &\bra{\Psi_\mathrm{BCS}(t)}i\partial_t\ket{\Psi_\mathrm{BCS}(t)} \label{derivativetGW}.
\end{align}
For details of derivation, see Gutzwiller approximation subsection in Methods.

Third, we derive the BdG Hamiltonian from the time-dependent Gutzwiller Hamiltonian in Eq.~\eqref{HG}. 
For order parameters, we consider two SU(2)-symmetric orders, i.e., the bond order amplitude $\chi_\tau(t)$ and the superconducting pairing amplitude $\Delta_\tau(t)$, 
\begin{align}
    \chi_{\tau}(t)=&\frac{1}{N}\sum_i \ev{\hat{c}_{i\uparrow}^\dagger \hat{c}_{i+\tau \uparrow}+\hat{c}_{i\downarrow}^\dagger \hat{c}_{i+\tau \downarrow}},\label{chi_update}\\
    \Delta_{\tau}(t)=&\frac{1}{N}\sum_i \ev{\hat{c}_{i\uparrow} \hat{c}_{i+\tau \downarrow}-\hat{c}_{i\downarrow}\hat{c}_{i+\tau \uparrow}},\label{Delta_update}
\end{align}
where $\tau=mx+ny$ represents a bond connecting two sites that are distant by $ma$ in the $x$ direction and $na$ in the $y$ direction, with $a$ being the lattice constant.
By applying the mean-field approximation to each term in Eq.~\eqref{HG} and performing a Fourier transformation, we arrive at the effective BdG Hamiltonian in the momentum-space representation as 
\begin{align}
  \hat{H}\smrm{BdG}(t)=&\sum_{\bm{k}}\left(\begin{array}{c}
    \hat{c}_{\bm{k}\uparrow}\\   \hat{c}_{-\bm{k}\downarrow}^\dagger
  \end{array}\right)^\dagger
  \mathcal{H}(\bm{k},t)\left(\begin{array}{c}
    \hat{c}_{\bm{k}\uparrow}\\   \hat{c}_{-\bm{k}\downarrow}^\dagger
  \end{array}\right), \label{Hbdg}
\end{align}
with
\begin{align}
  \mathcal{H}(\bm{k},t)&=\left(
    \begin{array}{cc}
      \varepsilon_{\bm{k}}(t) &F_{\bm{k}}(t)\\
      F_{\bm{k}}^*(t)&-\varepsilon_{-\bm{k}}(t)
    \end{array}
   \right).
\end{align}
Here, matrix elements are given by
\begin{align}
  \varepsilon_{\bm{k}}(t)=&-\frac{\delta }{\bar{f}}\sum_{m\tau} e^{-im\omega t} t_\tau^{(m)}e^{i\bm{k}\cdot \bm{R}_\tau}\nonumber \\
  &-\text{Re}\sum_{mn\tau \tau'}\frac{t_\tau^{(m-n)}t_{\tau'}^{(n)}e^{-im\omega t}}{U-n\omega}\Bigg\{ \frac{\delta f}{\bar{f}}e^{i\bm{k}\cdot (\bm{R}_\tau+\bm{R}_{\tau'})}(1-\delta_{\tau,-\tau'})\nonumber \\
  &+\frac{\chi_\tau(t) e^{i\bm{k}\cdot \bm{R}_{\tau'}}+\chi_{\tau'}(t) e^{i\bm{k}\cdot \bm{R}_{\tau}}}{4\bar{f}^2}[3(1-\delta)\delta_{\tau,-\tau'}+\delta(3-\delta)]\Bigg\}, \label{eps}\\
  F_{\bm{k}}(t)=&\frac{1}{2\bar{f}^2}\sum_{mn\tau \tau'}\frac{e^{-im\omega t}t_{\tau}^{(m-n)}t_{\tau'}^{(n)}+e^{im\omega t}t_{-\tau'}^{(-m+n)}t_{-\tau}^{(-n)}}{2(U-n\omega)}\nonumber \\
  &\times [3(1-\delta)\delta_{\tau,-\tau'}+ \delta(3+\delta)]\Delta_{\tau'}(t)\cos \bm{k}\cdot \bm{R}_\tau. \label{F}
\end{align}

Finally, we obtain the effective Lagrangian as $L_G \simeq \bra{\Psi_\mathrm{BCS}(t)}(i\partial_t -\hat{H}\smrm{BdG}(t))\ket{\Psi_\mathrm{BCS}(t)}$. 
By using the action principle with $\delta u_{\bm{k}}^*(t),\ \delta v_{\bm{k}}^*(t)$ as variants, 
we end up with the time-dependent Schr\"odinger equation,  
\begin{align}
  i\partial _t \vec{\psi}_{\bm{k}}(t)&= \mathcal{H}(\bm{k},t)\vec{\psi}_{\bm{k}}(t), \label{schrodingerEq} 
\end{align} 
where $\vec{\psi}_{\bm{k}}(t)=(v_{\bm{k}}(t),\ u_{\bm{k}}(t))^\mathrm{T}$. 
We solve this Schr\"odinger equation \eqref{schrodingerEq}, by computing the order parameters $\chi_\tau$ and $\Delta_\tau$ in Eqs.~\eqref{chi_update} and \eqref{Delta_update} consecutively.

\begin{figure*}[t]
\begin{center}
\includegraphics[width=\linewidth]{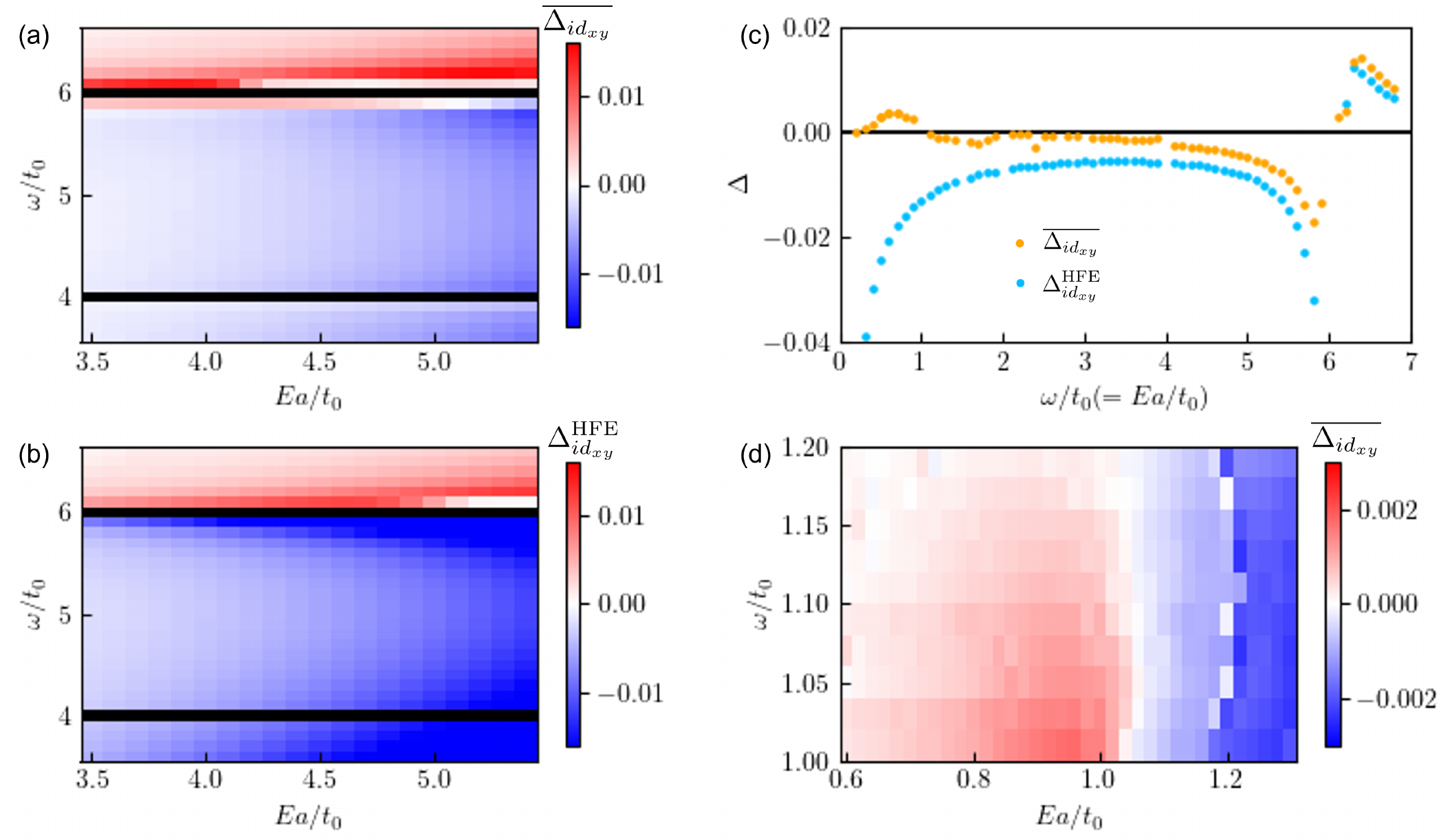}
\caption{Driving field dependence of the $id_{xy}$-wave component of the pairing amplitude.
(a) The time average of the $id_{xy}$-wave component of the pairing amplitude $\overline{\Delta_{id_{xy}}}$ in the steady state obtained from the time-dependent Gutzwiller simulation. 
We plot $\overline{\Delta_{id_{xy}}}$ as a function of the electric field $E$ and the frequency $\omega$ in the high-frequency region. 
(b) The $id_{xy}$-wave component of the pairing amplitude obtained from the high-frequency expansion $\Delta_{id_{xy}}^\mathrm{HFE}$. We plot $\Delta_{id_{xy}}^\mathrm{HFE}$
as a function $E$ and $\omega$ with the formula in Ref.~\cite{Kitamura2022}.
$\overline{\Delta_{id_{xy}}}$ and $\Delta_{id_{xy}}^\mathrm{HFE}$ show qualitatively similar behavior in the high-frequency region.
(c) Comparison of $\omega$ dependence of $\overline{\Delta_{id_{xy}}}$ and $\Delta_{id_{xy}}^\mathrm{HFE}$ with fixing $Ea/\omega =1$.
In the low-frequency region, the difference between $\overline{\Delta_{id_{xy}}}$ and $\Delta_{id_{xy}}^\mathrm{HFE}$ becomes significant, while they show a qualitative agreement in the high-frequency region.
$\overline{\Delta_{id_{xy}}}$ shows a sign change around $\omega=t_0$ that is not captured by HFE.
(d) $(E,\omega)$ dependence of $\overline{\Delta_{id_{xy}}}$ in the low-frequency region.
The sign change appears around $E=1.05t_0/a$, indicating a topological phase transition by increasing the electric field of the CPL. 
}
\label{fig:3}
\end{center}
\end{figure*}

\subsection*{Time evolution of superconducting pairing amplitudes}\label{sec:3}
In this section, we show the results of the time evolution generated by Eq.~\eqref{schrodingerEq}. 
We consider hopping amplitudes, bond order amplitudes, and superconducting pairing amplitudes up to next-nearest neighbors. 

Throughout this paper, we set the next-nearest-neighbor hopping $t_0'=-0.2t_0$, the onsite interaction $U=12t_0$, and the hole doping level $\delta=0.2$, with $t_0$ being the nearest-neighbor hopping amplitude.
As for the other doping levels and onsite interaction, see Figs. S1-S3 in Supplementary Note 1.

We show the time evolution of superconducting pairing amplitudes $\Delta_\tau$ for $\omega=5.8t_0$ in Figs.~\ref{fig:2}(b,c). 
The time profile of the driving field amplitude is shown in Fig.~\ref{fig:2}(a), and the convention for the superconducting pairing amplitudes is depicted in Fig.~\ref{fig:2}(g). 
As shown in Fig.~\ref{fig:2}(a), we increase the electric field amplitude linearly until $\omega t/2\pi=200$ and keep it constant after that.
We adopted this time profile of $E$ rather than just quenching the field, to connect the initial state to the dynamically-stabilized topological phase while mitigating damping oscillations in the order parameters.
We show four superconducting pairing amplitudes: $\Delta_{[1,0]},\Delta_{[1,1]},\Delta_{[0,1]}$ and $\Delta_{[-1,1]}$ in Fig.~\ref{fig:2}
(Other components can be obtained from the relationship $\Delta_{[x,y]}=\Delta_{[-x,-y]}$ for singlet pairing). 
We perform simulations with the initial state in the $d_{x^2-y^2}$-wave superconducting state which is the ground state of Eq.~\ref{Hbdg} in the absence of the field.
Namely, as shown in Figs.~\ref{fig:2}(b,c), we set positive 
$\Delta_{[1,0]}$ and negative $\Delta_{[0,1]}$ with the same magnitude while the others are zero, with which the $d_{x^2-y^2}$-wave component given by $(\mathrm{Re}\Delta_{[1,0]}-\mathrm{Re}\Delta_{[0,1]})/2$ has a nonzero value.

Once we turn on the electric field,
the phases of the superconducting pairing amplitudes start to rotate, as shown in Figs.~\ref{fig:2}(b,c). 
To make it easier to see the relative phases between different $\Delta_{\tau}$'s, we perform a phase rotation with the phase $\theta (t)$ of the pairing amplitude $\Delta_{[1,0]}$ 
(i.e., $\Delta_{[1,0]}(t)=|\Delta_{[1,0]}(t)|e^{i\theta (t)}$).
In Figs.~\ref{fig:2}(d,e), we plot the pairing amplitudes with the phase rotation, $\mathrm{Re}[\Delta_{\tau}e^{-i\theta }]$ and $\mathrm{Im}[\Delta_{\tau}e^{-i\theta }]$.
Figures~\ref{fig:2}(f,g) are the blowups of shaded areas of Figs.~\ref{fig:2}(d,e), which corresponds to the approximate steady state under the driving.
Here, the $d_{x^2-y^2}$-wave component is given by $(\mathrm{Re}[\Delta_{[1,0]}e^{-i\theta}]-\mathrm{Re}[\Delta_{[0,1]}e^{-i\theta}])/2$ and remains finite in the presence of the driving field.

Now let us look at the $id_{xy}$-wave component of the pairing amplitude.
We define the $id_{xy}$-wave component of superconducting pairing amplitudes as $\Delta_{id_{xy}}\equiv (\mathrm{Im}[\Delta_{[-1,1]}e^{-i\theta}]-\mathrm{Im}[\Delta_{[1,1]}e^{-i\theta}])/2$, which indeed belongs to the same irreducible reprenentation of $C_{4v}$ as the $d_{xy}$ pairing.
Figure~\ref{fig:2}(f) shows the time evolution of $\Delta_{id_{xy}}$, and
Fig.~\ref{fig:2}(h) is its blowup of the shaded area.
While $\Delta_{id_{xy}}$ is zero in the initial state without the CPL,
$\Delta_{id_{xy}}$ becomes finite once the CPL is applied, clearly indicating that the CPL irradiation induces  $id_{xy}$-wave component of the pairing amplitude leading to topological superconductivity.

Next, to study the magnitude of the $id_{xy}$-wave pairing amplitude in the steady state, we compute the time average of $\Delta_{id_{xy}}$ over $\omega t/2\pi \in [240,400]$, denoted as $\overline{\Delta_{id_{xy}}}$.
We show a color plot of $\overline{\Delta_{id_{xy}}}$ as a function of the driving amplitude $E$ and the frequency $\omega$ in Fig.~\ref{fig:3}(a). 
Note that for $\omega=4t_0,\ 6t_0$, Eqs.~\eqref{eps} and \eqref{F} contain divergent terms with the denominator $U-n\omega$ for $U=12t_0$, with which the effective Hamiltonian becomes ill-defined because of breaking down of the Schrieffer-Wolff transformation. 
We avoided those parameters in plotting Fig.~\ref{fig:3}(a).
For comparison, we show the $id_{xy}$-wave component $\Delta_{id_{xy}}^\mathrm{HFE}$ obtained from the high-frequency expansion of $\hat{H}\smrm{SW}$ in Fig.~\ref{fig:3}(b) 
[Specifically, we compute $\Delta_{id_{xy}}^\mathrm{HFE}\equiv (\mathrm{Im}\Delta_{[-1,1]}^\mathrm{HFE}-\mathrm{Im}\Delta_{[1,1]}^\mathrm{HFE})/2$, where $\Delta_{\tau}^\mathrm{HFE}$ is the superconducting pairing amplitude in the ground state of the Floquet Hamiltonian obtained by applying the HFE to Eq.~\ref{HSW}~\cite{Kitamura2022}. 
Figures~\ref{fig:3}(a,b) show that the $id_{xy}$-wave components obtained in the present approach are generally consistent with the HFE in the high-frequency region of $\omega>5t_0$. 

The present time-dependent Gutzwiller approach does not rely on the HFE and is also applicable to the low-frequency region in contrast to the HFE approach.
Here we fix $Ea/\omega =1$ and compare  $\overline{\Delta_{id_{xy}}}$ with $\Delta_{id_{xy}}^\mathrm{HFE}$ as functions of $\omega$ in Fig.~\ref{fig:3}(c). 
$\overline{\Delta_{id_{xy}}}$ and $\Delta_{id_{xy}}^\mathrm{HFE}$ show similar behaviors in the high-frequency region $\omega>5t_0$, while they are significantly different in the low-frequency region $\omega<2t_0$.
$\overline{\Delta_{id_{xy}}}$ even shows a sign change around $\omega \sim t_0$ as opposed to $\Delta_{id_{xy}}^\mathrm{HFE}$.
This topological phase transition can also be seen in Fig.~\ref{fig:3}(d), a color plot of $\overline{\Delta_{id_{xy}}}$ as a function of $E$ and $\omega$ in the low-frequency region. 
Specifically, the topological phase transition clearly occurs around $E=1.05t_0/a$. 
In this way, the present method can capture detailed behaviors of the Floquet topological superconductivity even in the low-frequency region where the HFE is not applicable.
In particular, topological phase transitions can be induced by changing the field strength of CPL in the low-frequency region.

\begin{figure*}[t]
\begin{center}
\includegraphics[width=\linewidth]{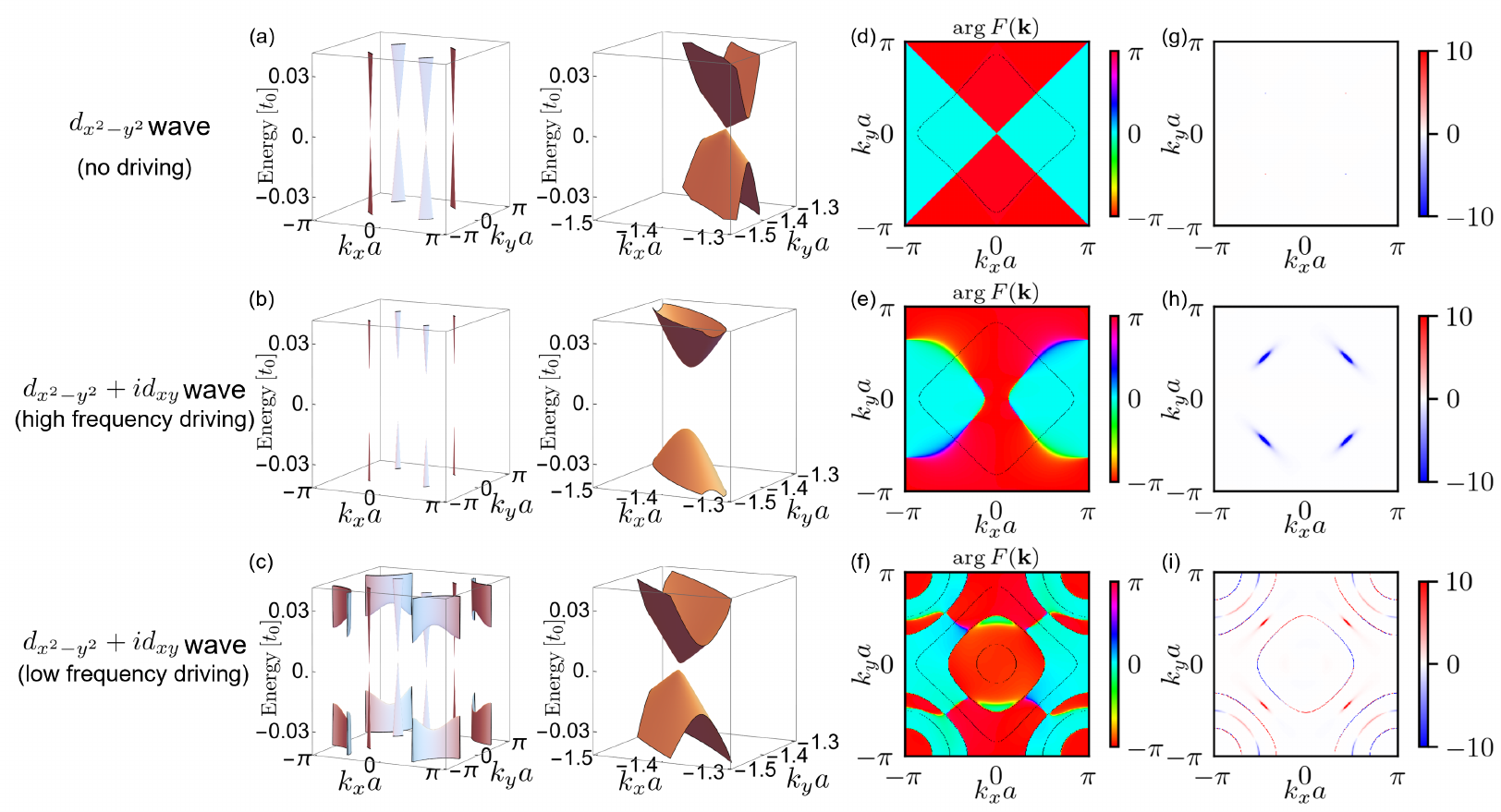}
\caption{Topological properties of Floquet energy bands for a $d$-wave superconductor without the CPL driving (a,d,g), with the CPL driving of high frequency (b,e,h), and with the CPL driving of low frequency (c,f,i). (a-c) Energy dispersion of Floquet bands with blowups around the point nodes. Without driving point nodes exist in the nodal line direction (a). Those nodes are gapped out when the CPL driving is applied (b,c). 
(d-f) The phase of gap functions $F(\bm{k})$ in the Floquet Hamiltonian.
The black curves indicate the $k$ points where the diagonal (normal) component of the Floquet Hamiltonian is zero.
Without driving (d), the gap function is real and has nodal lines. The black curve coincides with the Fermi surface in the normal state.
In the high-frequency driving case (e), the phase winds twice in the counterclockwise direction along the black curve. 
In the low frequency driving case (f), the phase winds twice in the clockwise direction along the black curve. Additional black curves arise from the folding of the Floquet bands.
(g-i) Berry curvatures of the Floquet bands with and without the CPL driving.
The driving frequency $\omega$ and the field amplitude $E$ are chosen as $\omega  =5.8t_0,\ E =5.4t_0/a$ for (b,e,h) and $\omega  =1.03t_0,\ E =0.96t_0/a$ for (c,f,i). 
}
\label{fig:bandGapCurv}
\end{center}
\end{figure*}

\subsection*{Floquet theory analysis}\label{sec:4}
The Floquet theory is a useful tool to describe periodically driven systems via an effective static Hamiltonian (Floquet Hamiltonian).
The topological properties of the steady state in the CPL driven system are characterized by the Chern number of the Floquet Hamiltonian. 
In this section, we show the Floquet bands and Chern numbers obtained in the present method.

\subsubsection*{Floquet Hamiltonian}
The Floquet Hamiltonian is defined as $\mathcal{H}_F(\bm{k})\equiv \frac{i}{T} \log \mathcal{T} \left\{ \exp\left[  -i \int_{t_0}^{t_0+T} \mathcal{H}(\bm{k}, t) dt\right] \right\}$ for a time-periodic Hamiltonian $\mathcal{H}(\bm{k}, t)$ of a period $T$. 
As can be seen from Figs.~\ref{fig:2}(b,c,h,i),  $\hat{H}\smrm{BdG}(t)$ is not time-periodic due to the phase rotation of $\Delta_{\tau}$  over a period, while $\Delta_\tau e^{-i\theta}$ becomes (approximately) time-periodic. 
In the following, we derive the Floquet Hamiltonian using this phase rotation for $\Delta_{\tau}$. 

In the present system, once the system arrives at the steady state, the superconducting pairing amplitudes satisfy
\begin{align}
  \Delta _{\tau}(t+T)=\Delta _{\tau}(t)e^{i\alpha},
\end{align}
as can be seen from Figs.~\ref{fig:2}(b,c).
We can remove this phase rotation by a gauge transformation $\vec{\psi}_{\bm{k}}'\equiv e^{i\epsilon t\tau_z}\vec{\psi}_{\bm{k}}$ with $-\epsilon T\equiv \alpha/2$, where $\vec{\psi}_{\bm{k}}'$ becomes time-periodic.
Here, $\tau_z$ is a Pauli $z$ matrix acting on the Nambu space.
Correspondingly, the Hamiltonian becomes time-periodic within this gauge and enables us to define the Floquet Hamiltonian as 
\begin{align}
  \mathcal{H}_F(\bm{k})\equiv \frac{i}{T}\log \left[e^{-i\frac{\alpha }{2}\tau_z}U(T,0)\right], \label{FloquetH}
\end{align}
where $U(T,0)$ is the original time evolution operator.
For detailed derivation, see Time-periodic Hamiltonian and Floquet Hamiltonian subsection in Methods.

\begin{figure*}[t]
\begin{center}
\includegraphics[width=0.8\linewidth]{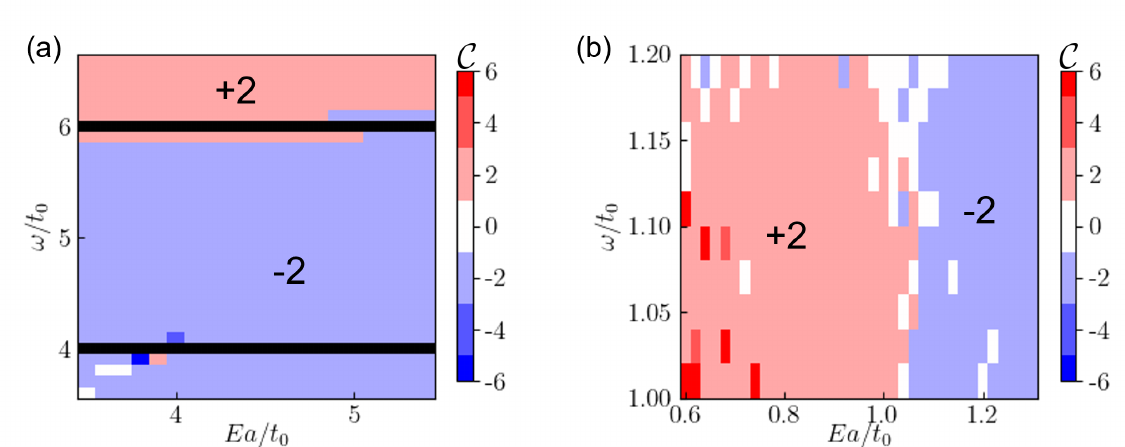}
\caption{Phase diagram of Floquet topological superconductors with the CPL driving.
The color plot of Chern number as a function of $(E,\omega)$ (a) in the high-frequency region and (b) in the low-frequency region.
The overall feature of the phase diagram is consistent with the sign of the $id_{xy}$-wave component of the pairing amplitude $\overline{\Delta_{id_{xy}}}$.
Jumps of the Chern number appear due to the folding of the Floquet bands in the low-frequency region.
}
\label{fig:phaseDiagram}
\end{center}
\end{figure*}

\subsubsection*{Floquet band and Chern number}\label{bandAndChern}
We compute the Floquet Hamiltonian for $\mathcal{H}(\bm{k}, t)$ in Eq.~\eqref{schrodingerEq} with the above method and obtain the associated Floquet bands. 
We then compute the Berry curvature and the Chern number for the Floquet bands by using the Fukui-Hatsugai-Suzuki method~\cite{Fukui2005}. 
Note that the results below do not take into account the occupation of the Floquet band.

Figure~\ref{fig:bandGapCurv} shows the obtained Floquet bands, the phases of gap functions $F(\bm{k})$ of the Floquet Hamiltonian, and the Berry curvature.
The results for the system without driving fields are shown in Figs.~\ref{fig:bandGapCurv}(a,d,g).
Figure~\ref{fig:bandGapCurv}(d) indicates that the gap function has the conventional $d_{x^2-y^2}$-wave symmetry with nodal lines. The black curve in Fig.~\ref{fig:bandGapCurv}(d) indicates the Fermi surface in the normal state. The point nodes appear at the crossing points of nodal lines and the Fermi surface as seen in Fig.~\ref{fig:bandGapCurv}(a).

Next, let us look at the results for the systems with the CPL drivings.
We study two cases of CPL driving.
One is in the high-frequency regime with
the driving frequency $\omega =5.8t_0$ and the field amplitude $E =5.4t_0/a$, shown in Figs.~\ref{fig:bandGapCurv}(b,e,h).
The other is in the low-frequency regime with $\omega =1.03t_0$ and $E=0.96t_0/a$, shown in Figs.~\ref{fig:bandGapCurv}(c,f,i).
With the CPL drivings, we find that the point nodes are gapped out in the Floquet band structures in both cases in Figs.~\ref{fig:bandGapCurv}(b,c).
We show the phase of gap functions obtained from the Floquet Hamiltonians with the CPL drivings in Figs.~\ref{fig:bandGapCurv}(e,f). 
We remark that the Floquet Hamiltonian $\mathcal{H}(\bm{k})$ and the gap function depend on the initial time $t_0$ of the time evolution operator $U(t_0+T,t_0)$ (while the eigenvalues are independent).
Although the gap function shown in Figs.~\ref{fig:bandGapCurv}(e,f) breaks the $C_4$ symmetry, the $C_4$-breaking component rotates with $t_0$, so that the Floquet state preserves $C_4$ symmetry in time average.

In these cases, the black curves correspond to the $k$ points where the diagonal (normal) component of the Floquet Hamiltonian is zero (Fermi energy in the equilibrium cases) or $\omega /2$ (Floquet Brillouin zone boundary).
In the high-frequency case ($\omega =5.8t_0,\ E =5.4t_0/a$), the black curve in Fig.~\ref{fig:bandGapCurv}(e) almost coincides with the Fermi surface in the equilibrium case [Fig.~\ref{fig:bandGapCurv}(d)].
Due to the emergence of the $id_{xy}$-wave component of the pairing amplitude,
the phase of the gap function rotates twice in the counterclockwise direction along the black curve [Fig.~\ref{fig:bandGapCurv}(e)].
As a consequence, a negative Berry curvature appears around the area where the point node was originally located [Fig.~\ref{fig:bandGapCurv}(h)]. 
In the low-frequency case ($\omega  =1.03t_0,\ E =0.96t_0/a$), Fig.~\ref{fig:bandGapCurv}(f) shows additional black curves in addition to the one corresponding to the Fermi surface in the equilibrium. 
These additional black curves correspond to band crossings arising from the folding of copies of Floquet bands (photon-dressed states) onto the Floquet Brillouin zone.
If we focus on the black curve from the original Fermi surface, the phase of $F(\bm{k})$ rotates twice in the clockwise direction along the black curve in Fig.~\ref{fig:bandGapCurv}(f).
Correspondingly, a positive Berry curvature appears around the gapped node in Fig.~\ref{fig:bandGapCurv}(i).
As can be seen from Figs.~\ref{fig:bandGapCurv}(h,i), the Berry curvature $B(\bm{k})$ satisfies the relation $B(\bm{k})=B(-\bm{k})$, which indicates that the time-reversal symmetry is broken while the inversion symmetry is preserved, as is consistent with the $d_{x^2-y^2} + id_{xy}$ pairing.

Furthermore, the sign of the Berry curvature coincides with the sign of the $id_{xy}$-wave component of the pairing amplitude $\overline{\Delta_{id_{xy}}}$ in Fig.~\ref{fig:3}.
This clearly indicates that the CPL induced Berry curvature originates from the gap opening at the point node with the $id_{xy}$-wave component $\overline{\Delta_{id_{xy}}}$.
Specifically, the electronic structure of the original point node is described by a gapless Dirac fermion. Once the mass gap is introduced by $\overline{\Delta_{id_{xy}}}$ to the Dirac fermion, 
the Berry curvature of the sign of the mass $\overline{\Delta_{id_{xy}}}$ emerges.

We show the phase diagram of Floquet topological superconductivity by computing Chern numbers of the Floquet bands in Fig.~\ref{fig:phaseDiagram}.
We show the phase diagrams for the high-frequency region in Fig.~\ref{fig:phaseDiagram}(a) and the low-frequency region in Fig.~\ref{fig:phaseDiagram}(b). 
The location of topological phases in Fig.~\ref{fig:phaseDiagram} is generally consistent with the sign of the $id_{xy}$-wave component of the pairing amplitude $\overline{\Delta_{id_{xy}}}$ in Fig.~\ref{fig:3}.
Again, we avoid $\omega=4t_0,\ 6t_0$ in plotting Fig.~\ref{fig:phaseDiagram} because of the ill-defined effective Hamiltonian with the divergent denominator $U-n\omega$ in Eqs.~\eqref{eps} and \eqref{F}.
We note that there are some regions with $C\neq \pm 2$ showing jumps of the Chern number in Fig.~\ref{fig:phaseDiagram}.
They are mainly due to the Berry curvature around the additional black curves in Fig.~\ref{fig:bandGapCurv}(i).
Since they appear from the folding of copies of Floquet bands, usually the occupation of energy states does not change abruptly across the energy gap
(See Occupation of Floquet band subsection in Methods for detail).
Namely, although those Floquet bands have large Chern numbers, the states in the both lower and upper bands around those minigaps are equally occupied and their contributions to the expectation value of the (thermal) Hall response in state $\ket{\Psi (t)}$ cancel out.
Therefore, Berry curvature around the original point node gives a dominant contribution to the (thermal) Hall response when the electron occupation is not far from the equilibrium one.

\section*{Discussion}\label{sec:5}
We have demonstrated the emergence of topological superconductivity even at low frequencies.
To discuss the experimental feasibility, let us turn our attention to evaluating the size of topological gaps and the required magnitude of electric fields and frequencies.
The low-frequency region shown in Fig.~\ref{fig:phaseDiagram}(b) corresponds to $\hbar \omega \simeq \SIrange{0.4}{0.5}{\eV}$, $E\simeq \SIrange{8}{17}{\mega\V\per\cm}$ for the typical cuprates with $t_0 \simeq \SI{0.4}{\eV}$, $a\simeq \SI{3}{\AA}$.
One of the advantages of the present time-dependent Gutzwiller approach in the low-frequency regime is that it does not require the large electric fields as large as \SI{100}{\mega\V\per\cm} as mentioned in previous research~\cite{Takasan2017,Kitamura2022}, due to the effective coupling through the vector potential $A\propto E/\omega$.
The size of the topological gap appearing in Fig.~\ref{fig:bandGapCurv}(c) is approximately $0.03$ times that of the original $d_{x^2-y^2}$-wave superconducting gap (the gap function at the Fermi surface in the antinodal direction).
Therefore, assuming that $T_c$ of cuprate is \SI{89}{\K}~\cite{Ren2012}, experimental observation of the topological gap can be achieved at around \SI{3}{\K}.
While the magnitude of the topological gap is smaller than that indicated in Ref.~\cite{Kitamura2022}, the gap described in that reference scales as $\propto O(E^4)$.
At $E\sim \SI{10}{\mega\V\per\cm}$, the magnitude of the topological gap obtained by the present method is overwhelmingly larger.
In addition, as for the impurity effect, since the present $d+id$-wave superconductivity is based on the $d_{x^2-y^2}$-wave superconductivity in the undriven cuprates, we consider that the suppression of the topological superconductivity by the impurities will be qualitatively the same as that for the original $d$-wave superconductivity in cuprate superconductors.

Several experimental methods are considered for probing the Floquet topological superconductivity demonstrated in this paper.
The first is the measurement of the optical conductivity which directly observes the topological gap.
Several studies have already investigated the optical conductivity of cuprate superconductors~\cite{Wang1998,Tajima_2016}.
The irradiation of CPL opens the gap at the Dirac nodes in cuprate superconductors.
Therefore, during CPL irradiation, optical absorption at frequencies below the topological gap is expected to be suppressed.
The second is to detect the Higgs mode associated with the $id_{xy}$-wave component of the superconducting pairing amplitudes.
The $d_{x^2-y^2}$-wave superconductivity transforms into $d_{x^2-y^2}+id_{xy}$-wave superconductivity under CPL irradiation.
The Higgs mode of $d$-wave superconductivity in cuprate superconductors was investigated using terahertz pump and optical probe techniques~\cite{Katsumi2018,Shimano2020}.
Thus, observing the Higgs mode of the additional $id_{xy}$-wave component would provide evidence of Floquet topological superconductivity.
Another approach involves measuring the Hall response.
CPL irradiation breaks time-reversal symmetry and induces a finite Berry curvature.
The realization of a QAHI in graphene through CPL irradiation was recently observed using a laser-triggered photoconductive switch~\cite{McIver2020}. 
Similar methods may capture the Hall response in the present system.

\section*{Methods}
\subsection*{Schrieffer-Wolff transformation}\label{SWtrans}
In this section, we derive Eq.~\eqref{HSW}.
For clarity, we rewrite each term in $\hat{H}\smrm{Hub}$ as 
\begin{align}
-\sum_{ij\sigma }t_{ij}e^ {-i\bm{A}(t)\cdot \bm{R}_{ij}}\hat{c}^\dagger_{i\sigma}\hat{c}_{j\sigma}&=
   -[\hat{T}_{-1}(t)+\hat{T}_{0}(t)+\hat{T}_{+1}(t)],\\
  U\sum_i \hat{n}_{i\uparrow} \hat{n}_{i\downarrow}&= U\hat{D},
\end{align}
 where $\hat{T}_{d}(t)$ is the kinetic energy operator that changes the double occupancy by $d$: $[\hat{D},\hat{T}_{d}(t)]=d\hat{T}_{d}(t)$ with $\hat{D}$ being the double occupancy operator.
 The explicit form of $\hat{T}_{d}(t)$ is given by
\begin{align}
  & \hat{T}_{0}(t)=\sum_{i j \sigma} t_{i j}e^ {-i\bm{A}\cdot \bm{R}_{ij}}[(1-\hat{n}_{i \bar{\sigma}}) \hat{c}_{i \sigma}^{\dagger} \hat{c}_{j \sigma}(1-\hat{n}_{j \bar{\sigma}})+\hat{n}_{i \bar{\sigma}} \hat{c}_{i \sigma}^{\dagger} \hat{c}_{j \sigma}\hat{n}_{j \bar{\sigma}}], \label{T0t}\\
  & \hat{T}_{-1}(t)=\sum_{i j \sigma} t_{i j}e^ {-i\bm{A}\cdot \bm{R}_{ij}} (1-\hat{n}_{i \bar{\sigma}})\hat{c}_{i \sigma}^{\dagger} \hat{c}_{j \sigma}\hat{n}_{j \bar{\sigma}} =\hat{T}_{+1}^{\dagger}(t),  \label{T1t}
\end{align}  
where $\bar{\uparrow}\equiv \downarrow, \bar{\downarrow}\equiv \uparrow$. 

In order to obtain the transformed Hamiltonian $\hat{H}\smrm{SW}(t)=e^{i\hat{S}(t)}\hat{H}_\mathrm{Hub}(t)e^{-i\hat{S}(t)}-e^{i\hat{S}(t)}(i\partial_te^{-i\hat{S}(t)})$ with no charge excitations (i.e. $[\hat{D},\hat{H}\smrm{SW}(t)]=0$), 
let us determine $\hat{S}(t)$ order by order in terms of the hopping amplitude.
We can write down the first-order term of the transformed Hamiltonian as
\begin{align}  
\hat{H}\smrm{SW}^{(1)}=-[\hat{T}_{-1}+\hat{T}_{0}+\hat{T}_{+1}]+[i\hat{S}^{(1)},U\hat{D}]-\partial_t\hat{S}^{(1)},
\end{align}  
where the superscript $(n)$ for $\hat{S}(t)$ and $\hat{H}\smrm{SW}(t)$ denotes the order of the hopping amplitude, and we have chosen $\hat{S}^{(0)}(t)=0$. 
In order to satisfy $[\hat{D},\hat{H}\smrm{SW}^{(1)}(t)]=0$, $S^{(1)}(t)=S_{+1}^{(1)}(t)+S_{-1}^{(1)}(t)$ have to be chosen such that 
\begin{align}  
-\hat{T}_{\pm1}\mp iU\hat{S}_{\pm1}^{(1)}-\partial_t\hat{S}_{\pm1}^{(1)}=0,\label{eq:S1}
\end{align}  
with which the transformed Hamiltonian is given as $\hat{H}\smrm{SW}^{(1)}(t)=-\hat{T}_{0}(t)$.
The solution of the above operator equation for $\hat{S}_{\pm1}^{(1)}(t)$ is given as the same form as $\hat{T}_{\pm1}(t)$ but with $s_{ij,\pm1}(t)$ instead of $t_{ij}e^ {-i\bm{A}(t)\cdot \bm{R}_{ij}}$, where 
\begin{align}
s_{ij,\pm1}(t)=\pm \sum_m \frac{it_{ij}^{(m)}e^{-im\omega t}}{U\mp m\omega}\label{eq:S1periodic}
\end{align}
for the time-periodic hopping amplitude.

In a similar manner, the second-order term can be obtained as
\begin{align}  
\hat{H}\smrm{SW}^{(2)}(t)=
-\frac{1}{2}[i\hat{S}_{+1}^{(1)}(t),\hat{T}_{-1}(t)]+h.c.
\end{align}
We obtain Eq.~\eqref{HSW} by inserting Eqs.~\eqref{T0t}, \eqref{T1t}, and \eqref{eq:S1periodic} to the above expression. 
In the present method, we perform the Schrieffer-Wolff transformation up to the second-order of hopping.
This generates three-site term which induces topological superconductivity. 
Moreover, if terms of fourth order in Schrieffer-Wolff transformation are considered, scalar spin chirality terms scalar spin chirality terms develop, which also induce topological superconductivity~\cite{Kitamura2022}.

We note that the expression for $\hat{S}_{\pm1}^{(1)}(t)$ should be slightly modified when the field amplitude $\bm{A}_0$ becomes time-dependent~\cite{Kitamura2016,Eckstein2017}, while the correction is negligible when the change of the amplitude is slow enough.
The generalized expression for Eq.~\eqref{eq:S1periodic} is given as
\begin{align}  
s_{ij,\pm1}(t)=-\sum_m\int_{-\infty}^tdt^\prime t_{ij}^{(m)}(\bm{A}_0(t^\prime))e^{\mp i(U\mp i0^+)(t-t^\prime)-im\omega t^\prime}.
\end{align}
By expanding $t_{ij}^{(m)}$ in the Taylor series around $t^\prime=t$, we obtain 
\begin{align}  
s_{ij,\pm1}(t)=\sum_m \left[\pm\frac{it_{ij}^{(m)}(\bm{A}_0(t))}{U\mp m\omega}-\frac{\dot{\bm{A}}_0\cdot\partial_{\bm{A}_0}t_{ij}^{(m)}}{( U\mp m\omega)^2}+\dots\right]e^{-im\omega t}.
\end{align}
The correction term is smaller by a (dimensionless) factor of $\sim|\dot{\bm{A}}_0|a/U$, and is indeed negligible for the present calculation.

\subsection*{Gutzwiller approximation}\label{GWappro}
In this section, we describe the detailed derivation of Eqs.~\eqref{HG} and \eqref{derivativetGW}.
First, we evaluate the denominator of Eq.~\eqref{HSWtoHG} by site diagonal expectation values, which gives~\cite{Ogata2003,Ogata2008}
\begin{align}
    \bra{\Psi_\mathrm{BCS}}\hat{P}_G\hat{P}_G\ket{\Psi_\mathrm{BCS}}\simeq&\frac{N!}{(N\delta )!(Nf)!(Nf)!}(\bar{f}\bar{f})^{N\delta}(f\bar{f})^{Nf}(f\bar{f})^{Nf}\nonumber \\
    \sim &\frac{(\bar{f}\bar{f})^{N\delta}(f\bar{f})^{Nf}(f\bar{f})^{Nf}}{\delta ^{N\delta }f^{Nf}f^{Nf}}=(\bar{f}^2\delta ^{-1})^{N\delta} \bar{f}^{2Nf},
\end{align}
where $\delta=1-\frac{1}{N}\sum_i \ev{\hat{n}_{i\uparrow}+\hat{n}_{i\downarrow}}$ is the hole doping rate, $f=(1-\delta)/2$ is the probability to have a singly-occupied site in the Gutzwiller wave function $\hat{P}_G\ket{\Psi_\mathrm{BCS}}$, and $\bar{f}=1-f$. 
Here we have used $n! \sim (n/e)^n$ on the second line. 
We evaluate the expected value of each term in Eq.~\eqref{HSW} in a similar manner. For the hopping term, we obtain
\begin{align}
  &\ev{\hat{P}_G \hat{c}_{i\sigma}^\dagger \hat{c}_{j\sigma}\hat{P}_G}\nonumber \\
  \sim &\ev{(1-\hat{n}_{i\bar{\sigma}})\hat{c}_{i\sigma}^\dagger \hat{c}_{j\sigma}(1-\hat{n}_{j\hat{\sigma}})}\frac{(\bar{f}\bar{f})^{N\delta -1}(f\bar{f})^{Nf-1}(f\bar{f})^{Nf}}{\delta ^{N\delta -1}f^{Nf-1}f^{Nf}}\nonumber \\
=&\frac{\delta }{\bar{f}} \ev{\hat{c}_{i\sigma}^\dagger \hat{c}_{j\sigma}}\ev{\hat{P}_G},
\end{align}
while for the Heisenberg interaction we have
\begin{align}
    \ev{\hat{P}_G\hat{\bm{S}}_i\cdot \hat{\bm{S}}_j\hat{P}_G}\sim \frac{1}{\bar{f}^2}\ev{\hat{\bm{S}}_i\cdot \hat{\bm{S}}_j} \ev{\hat{P}_G}.
\end{align}
The spin part of the three-site term, $\hat{P}_G\left[(\hat{c}_{i\sigma}^\dagger \bm{\sigma}_{\sigma \sigma '}\hat{c}_{k\sigma '})\cdot \hat{\bm{S}}_j\right]\hat{P}_G$ with
$i\neq k$ is evaluated as
\begin{align}
    \ev{\hat{P}_G(\hat{c}_{i\sigma}^\dagger \bm{\sigma}_{\sigma \sigma '}\hat{c}_{k\sigma '})\cdot \hat{\bm{S}}_j\hat{P}_G}\sim \frac{\delta}{\bar{f}^2}\ev{(\hat{c}_{i\sigma}^\dagger \bm{\sigma}_{\sigma \sigma '}\hat{c}_{k\sigma '})\cdot \hat{\bm{S}}_j} \ev{\hat{P}_G}.
\end{align}
As for the density operator, there is arbitrariness in whether to renormalize this as an operator or to replace it as the density at each site. 
In the present case, in which $\delta$ has a finite value, the variational Monte Carlo calculation shows that the density-density term has a few effects in contrast to the half-filling case, which allows us to replace $\ev{\hat{P}_G\hat{n}_{i}\hat{P}_G}$ with $1-\delta$~\cite{Yokoyama1996}. 
We thus discard the second order fluctuations $(\hat{n}_{i\sigma}-f)(1-\hat{n}_{i\bar{\sigma}}-\bar{f})$ around the expected value of the density in the Hamiltonian here, like $\hat{P}_G\hat{n}_{i}\hat{P}_G \simeq \hat{n}_i\delta +2f^2$ with $\hat{n}_i=\hat{n}_{i\uparrow}+\hat{n}_{i\downarrow}$~\cite{Kitamura2022}. 
Then the remaining terms in the Hamiltonian can be evaluated as
\begin{gather}
    \ev{\hat{P}_G\hat{n}_i\hat{n}_j\hat{P}_G}\simeq \frac{\delta^2}{\bar f^2}\ev{\hat{n}_i\hat{n}_j}\ev{\hat{P}_G}+\frac{2\delta f^2}{\bar f^2}\ev{\hat{n}_i+\hat{n}_j}\ev{\hat{P}_G}+const.,\\
    \ev{\hat{P}_G\hat{c}_{i\sigma}^\dagger \hat{c}_{k\sigma}n_j\hat{P}_G}\simeq \frac{\delta^2}{\bar f^2}\ev{\hat{c}_{i\sigma}^\dagger \hat{c}_{k\sigma}\hat{n}_j}\ev{\hat{P}_G}+\frac{2\delta f^2}{\bar f^2}\ev{\hat{c}_{i\sigma}^\dagger \hat{c}_{k\sigma}}\ev{\hat{P}_G}
\end{gather}
with $i\neq k$. In this way, we obtain the effective Hamiltonian Eq~\eqref{HG}.

Evaluation of the first term of Eq.~\eqref{LG} incurs additional calculation of the derivative of the variational parameters.
Using the relation
\begin{align}
&(\dot{u}_{\bm{k}}+\dot{v}_{\bm{k}}\hat{c}_{\bm{k}\uparrow}^\dagger \hat{c}_{-\bm{k}\downarrow}^\dagger) \ket{0}\nonumber\\
&=\left(\frac{\dot{u}_{\bm{k}}}{u_{\bm{k}}}\hat{c}_{\bm{k}\uparrow}\hat{c}_{\bm{k}\uparrow}^\dagger+\frac{\dot{v}_{\bm{k}}}{v_{\bm{k}}}\hat{c}_{\bm{k}\uparrow}^\dagger \hat{c}_{\bm{k}\uparrow}\right)(u_{\bm{k}}+v_{\bm{k}}\hat{c}_{\bm{k}\uparrow}^\dagger \hat{c}_{-\bm{k}\downarrow}^\dagger)\ket{0},
\end{align}
we can rewrite the time derivative in an operator form as
\begin{align}
\partial_t \ket{\Psi_\mathrm{BCS}}=&\sum_{\bm{k}'}(\dot{u}_{\bm{k}'}+\dot{v}_{\bm{k}'}\hat{c}_{\bm{k}' \uparrow}^\dagger \hat{c}_{-\bm{k}' \downarrow}^\dagger) \prod_{\bm{k}\neq \bm{k}'}(u_{\bm{k}}+v_{\bm{k}}\hat{c}_{\bm{k}\uparrow}^\dagger \hat{c}_{-\bm{k}\downarrow}^\dagger)\ket{0}\nonumber \\
=&\sum_{\bm{k}}\left[\frac{\dot{u}_{\bm{k}}}{u_{\bm{k}}}+\left(\frac{\dot{v}_{\bm{k}}}{v_{\bm{k}}}-\frac{\dot{u}_{\bm{k}}}{u_{\bm{k}}}\right)\sum_{ij}\hat{c}_{i\uparrow}^\dagger \hat{c}_{j\uparrow}e^{-i\bm{k}\cdot (\bm{R}_i -\bm{R}_j)}\right]\ket{\Psi_\mathrm{BCS}},
\end{align}
with which the first term of Eq.~\eqref{LG} is written as
\begin{align}
  &\frac{\bra{\Psi_\mathrm{BCS}}\hat{P}_Gi\partial_t \hat{P}_G\ket{\Psi_\mathrm{BCS}}}{\bra{\Psi_\mathrm{BCS}}\hat{P}_G\hat{P}_G\ket{\Psi_\mathrm{BCS}}}\nonumber \\=&i\sum_{\bm{k}}\left[\frac{\dot{u}_{\bm{k}}}{u_{\bm{k}}}+\left(\frac{\dot{v}_{\bm{k}}}{v_{\bm{k}}}-\frac{\dot{u}_{\bm{k}}}{u_{\bm{k}}}\right)\left\{\sum_{i\neq j}\frac{\ev{\hat{P}_G\hat{c}_{i\uparrow}^\dagger \hat{c}_{j\uparrow}}}{\ev{\hat{P}_G}}e^{i\bm{k}\cdot \bm{R}_{ij}}+\sum_i \frac{\ev{\hat{P}_G\hat{n}_{i\uparrow}}}{\ev{\hat{P}_G}}\right\}\right].
\end{align} 
The expectation value $\ev{\hat{P}_G\hat{c}_{i\uparrow}^\dagger \hat{c}_{j\uparrow}}/\ev{\hat{P}_G}$  can be evaluated within the Gutzwiller approximation as
\begin{align}
  \frac{\ev{\hat{P}_G\hat{c}_{i\uparrow}^\dagger \hat{c}_{j\uparrow}}}{\ev{\hat{P}_G}}\simeq &\frac{(\bar{f}^2/\delta)^{N\delta }\bar{f}^{2Nf-2}\ev{(1-\hat{n}_{i\downarrow})\hat{c}_{i\uparrow}^\dagger \hat{c}_{j\uparrow} \hat{n}_{j\downarrow}}}{(\bar{f}^2/\delta)^{N\delta }\bar{f}^{2Nf}}\nonumber \\
  &+\frac{(\bar{f}^2/\delta)^{N\delta -1}\bar{f}^{2Nf-1}\ev{(1-\hat{n}_{i\downarrow})\hat{c}_{i\uparrow}^\dagger \hat{c}_{j\uparrow} (1-\hat{n}_{j\downarrow})}}{(\bar{f}^2/\delta)^{N\delta }\bar{f}^{2Nf}} \nonumber \\
  =&\left(\frac{f}{\bar{f}}+\frac{\delta}{\bar{f}}\right)\ev{\hat{c}_{i\uparrow}^\dagger \hat{c}_{j\uparrow}}=\ev{\hat{c}_{i\uparrow}^\dagger \hat{c}_{j\uparrow}}.
\end{align}
Note that both doubly-occupied and singly-occupied configurations at site $j$ are allowed as the initial configuration due to the absence of the projection operator.
The other expectation value can also be evaluated in the same way as $\ev{\hat{P}_G \hat{n}_{i\uparrow}}/\ev{\hat{P}_G}\simeq \ev{\hat{n}_{i\uparrow}}$, with which Eq.~\eqref{derivativetGW} follows.

\subsection*{Time-periodic Hamiltonian and Floquet Hamiltonian}\label{AppFloquetH}
In this section, we describe how to obtain the time-periodic BdG Hamiltonian and the associated Floquet Hamiltonian \eqref{FloquetH}.
The superconducting pairing amplitude of each bond $\tau$ can be computed from the gap equation,
\begin{align}
  \Delta _{\tau}(t)=2\sum_{\bm{k}} \vec{\psi}_{\bm{k}}^\dagger\left(\begin{array}{cc}
    0&0\\
    -1&0
  \end{array}\right)\vec{\psi}_{\bm{k}}\cos \bm{k}\cdot \bm{R}_\tau .
\end{align}
This implies that the pairing amplitude is transformed as $\Delta_{\tau}(t)\to\Delta_{\tau}^\prime(t)=\Delta_{\tau}(t)e^{2i\epsilon t}$
under the unitary transformation $\vec{\psi}_{\bm{k}}\to\vec{\psi}_{\bm{k}}'\equiv e^{i\epsilon t\tau_z}\vec{\psi}_{\bm{k}}$, 
which keeps the relative phase of the pairing amplitudes intact.
When the pairing amplitudes satisfy $\Delta_\tau(t+T)=\Delta_\tau(t)e^{i\alpha}$, the transformed amplitude becomes time-periodic 
\begin{align}
\Delta_{\tau}^\prime(t+T)=\Delta_{\tau}^\prime(t)e^{i\alpha+2i\epsilon T}=\Delta_{\tau}^\prime(t)
\end{align}
when $-\epsilon T=\alpha/2$. Namely, under this condition the transformed BdG Hamiltonian $\mathcal{H}'(\bm{k},t)\equiv e^{i\epsilon t\tau_z}\mathcal{H}(\bm{k},t)e^{-i\epsilon t\tau_z}-\epsilon \tau_z$ becomes time-periodic, and we can define the corresponding Floquet Hamiltonian, 
\begin{align}
\mathcal{H}_F(\bm{k},t_0)\equiv \frac{i}{T}\log \left[\mathcal{T} \exp(-i\int_{t_0}^{t_0+T}\mathcal{H}'(\bm{k},t)dt)\right],
\end{align}
where the time evolution of its eigenstates are expressed as the time-periodic wave function multiplied by a plane wave factor.
Using the relation for the transformed time evolution operator, $U^\prime(t,t^\prime)=e^{i\epsilon t\sigma_z}U(t,t')e^{-i\epsilon t'\sigma_z}$, we can compute the Floquet Hamiltonian as
\begin{align}
  \mathcal{H}_F(\bm{k},\ t_0):&=\frac{i}{T} \log U'(t_0+T,t_0)\\
  U'(t_0+T,t_0):&=e^{i\epsilon (t_0+T)\sigma _z}U(t_0+T,t_0)e^{-i\epsilon t_0\sigma _z}.
\end{align}
By setting $t_0=0$, we arrive at Eq.~\eqref{FloquetH} with 
$\vec{\psi}_{\bm{k}}'(t_0)=\vec{\psi}_{\bm{k}}(t_0)$.

\subsection*{Occupation of Floquet band}\label{Occupation}
\begin{figure}[t]
  \begin{center}
    \includegraphics[width=\linewidth]{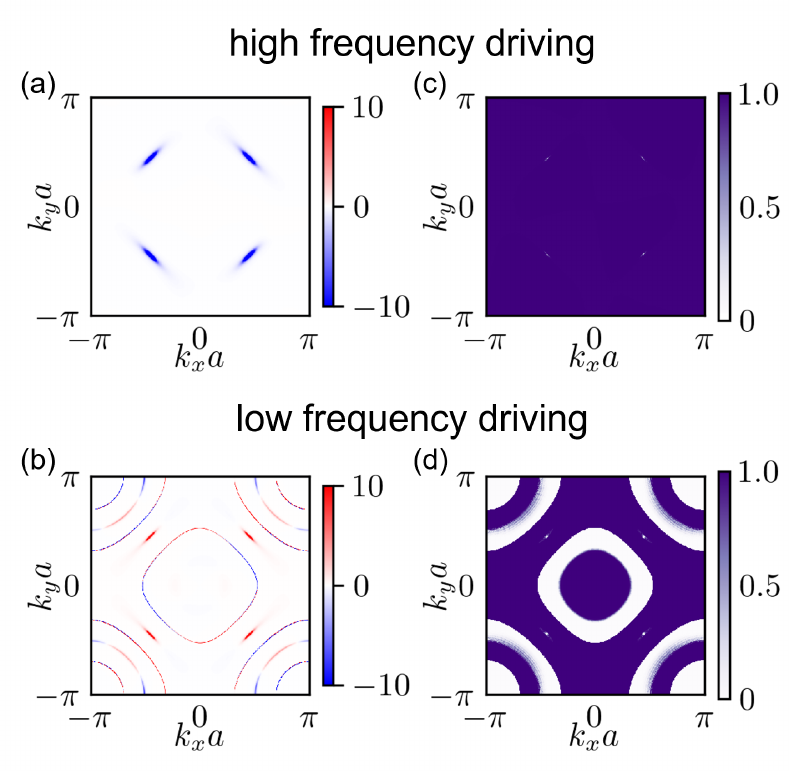}
    \caption{Occupation probabilities of the lower Floquet band in the Brillouin zone after the simulation
      with the CPL driving of (c) high frequency ($\omega  =5.8t_0,\ E =5.4t_0/a$) and (d) low frequency ($\omega  =1.03t_0,\ E =0.96t_0/a$).
      The situations correspond to Figs.~\ref{fig:bandGapCurv}(b,e,h) and Figs.~\ref{fig:bandGapCurv}(c,f,i), respectively.
      For comparison, Figs.~\ref{fig:bandGapCurv}(h,i) are reproduced as (a,b).
    }
    \label{fig:distri}
  \end{center}
\end{figure}
Figures~\ref{fig:distri}(c,d) show the occupation probabilities of the lower Floquet band after the time evolution with the CPL driving.
For comparison, we replot the Berry curvature in Figs.~\ref{fig:bandGapCurv}(h,i) as Figs.~\ref{fig:distri}(a,b).
In the high-frequency case ($\omega =5.8t_0,\ E =5.4t_0/a$), the lower Floquet band is almost fully occupied except for the regions where the point nodes were originally located.
The lack of occupancy in these regions is attributed to the choice of our initial state in which the energy gap closes.
These unoccupied regions should be filled when the scattering processes neglected in the present mean-field treatment are taken into account.
In the low-frequency case ($\omega =5.8t_0,\ E =5.4t_0/a$), other unoccupied regions arise due to the foldings of copies of the Floquet band.
When the external electric field is turned on slowly as in the present case, the occupation of Floquet bands is almost determined by the original distribution function in the absence of the external field. 
In those cases, the boundaries of the occupied and unoccupied areas of the Floquet bands appear at the position of band crossings arising from band folding in the Floquet Brillouin zone as seen in Fig.~\ref{fig:distri}.
In general, the occupation of Floquet bands strongly depends on the time profile of the electric field and presence or absence of scatterings. 

\textbf{Data Availability:} 
Data are available upon reasonable request.

\textbf{Code Availability:}
Code is available upon reasonable request.

\bibliographystyle{naturemag}
\bibliography{references}

\acknowledgements
We thank Ryo Shimano, Takashi Oka and Masamitsu Hayashi for fruitful discussions.
This work was supported by 
JSPS KAKENHI Grant 23H01119, 23K17665 (T.M.) and 20K14407 (S.K.), 
JST CREST (Grant No. JPMJCR19T3) (T.M., S.K.).

\textbf{Author Contributions:}
S.K. and T.M. conceived the project. T.A. has performed the theoretical formulation and numerical calculations. All authors wrote the manuscript.

\textbf{Competing Interests:}
The authors declare no competing interests.

\end{document}